\documentclass[twocolumn,amsmath,amssymb,fixfloats]{revtex4}

\usepackage[english]{babel}
\usepackage{amsmath}
\usepackage{amssymb}
\usepackage{graphicx}

\begin{document}

\title{Coupled dynamics of atoms and radiation pressure driven interferometers}

\author{D. Meiser}
\author{P. Meystre}

\affiliation{Department of Physics, The University of Arizona,
1118 E. 4th Street, Tucson, AZ 85721 }

\begin{abstract}
We consider the motion of the end mirror of a cavity in whose standing wave
mode pattern atoms are trapped. The atoms and the light field strongly couple
to each other because the atoms form a distributed Bragg mirror with
a reflectivity that can be fairly high. We analyze how the dipole potential in
which the atoms move is modified due to this backaction of the atoms. We show
that the position of the atoms can become bistable. These results are of a more
general nature and can be applied to any situation where atoms are trapped in
an optical lattice inside a cavity and where the backaction of the atoms on the
light field cannot be neglected. We analyze the dynamics of the coupled
system in the adiabatic limit where the light field adjusts to the position
of the atoms and the light field instantaneously and where the atoms move much
faster than the mirror. We calculate the side band spectrum of the light
transmitted through the cavity and show that these spectra can be used to
detect the coupled motion of the atoms and the mirror.
\end{abstract}
\maketitle

\section{\label{introduction} Introduction}

In recent years the coupling of the light field inside an optical
resonator to the mechanical motion of the end mirrors
\cite{Dorsel:OB_radiation_pressure,Pierre:Theo_RP_driven_interferometer,Solimeno:oscillatingmirrors}
has received renewed attention due to several developments. The
improved capabilities to micromachine sub-micrometer sized
mechanical structures with well defined mechanical and optical
properties enables the creation of moveable mirrors with small
damping whose motion is strongly affected by the radiation
pressure due to the light field inside an optical cavity.  The
coupled nonlinear dynamics found in those systems is very rich
\cite{Marquardt:Multistability,Stambaugh:switching_micro_oscillator}
and is relevant for systems as large as gravitation wave antennas
\cite{Meers:instabilities,Pierro:rpchaos,Aguirregabiria:instabilities}
down to micromachined cavities \cite{Marquardt:Multistability}.
The radiation pressure force can be used to isolate the mechanical
motion of the end mirror from the seismic background
\cite{Pierre:Theo_RP_driven_interferometer}. Recently, schemes for
active cooling of mirrors with radiation pressure have been
proposed \cite{Wilson_Rae:Laser_cooling_resonator} and implemented
\cite{Cohadon:RP_cooling,Hoehberger:cavity_cooling_microlever,Vitali:optomechanicalcooling}.
It is expected that with these methods cooling of the mirror
motion to its quantum mechanical ground state will be possible and
that the quantized motion of a macroscopic object, including
non-classical states of motion and squeezing, can be studied this
way
\cite{Jacobson:Quantum_noise_position_measurement,Tittonen:interferometric_position_measurement,Courty:quantumlimits}.
Also, the mirror could become entangled with the light field
\cite{Marshall:Mirror_quantum_superpositions}, and questions
related to position measurements near the quantum limit can be
addressed. In the classical regime the mirror motion can exhibit
instabilities, self-sustained oscillations
\cite{Marquardt:Multistability} and chaos due to the retardation
of the electromagnetic field in certain cases.

From a theoretical point of view, radiation pressure driven
cavities offer many challenges and opportunities. These systems
provide excellent testing grounds for theories of systems with
delay \cite{Bel:delayed_systems,Aguirregabiria:instabilities}.
Furthermore quantizing the light field is a nontrivial question in
principle for time-dependent boundary conditions
\cite{Law:Rubber_cavity_quantization}.

\begin{figure}
\includegraphics{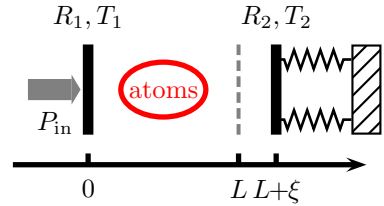}
\caption{(Color online) Schematic of atoms in a cavity with a moveable mirror.}
\label{setup}
\end{figure}

In this paper we extend the study of radiation pressure driven
interferometers to the case where atoms are trapped in the optical
dipole potential generated by the standing-wave light field inside
the cavity, see Fig. \ref{setup}. In this system the atoms, light
field and mirror interact with each other on an equal footing in
the most general case. We derive the equations governing this
system from first principles. Treating the motion of the mirror as
well as the light field classically, we show that in many cases of
interest the motion of the atoms can also be described
classically. We restrict ourselves to those situations in this
paper. Our key result is that, surprisingly if one considers the
complexity of the problem, it is possible to develop an effective
theory that captures many of the aspects of the system and is
amenable to a fairly simple analysis under moderate assumptions.
The key observation is that the effect of the atoms on the light
field can be represented by an effective mirror whose reflection
and transmission coefficient can be calculated from the
self-consistently determined atomic density distribution. As we
will discuss the atoms naturally arrange themselves in a
configuration corresponding to a Bragg mirror of maximum
reflectivity for a given number of atoms
\cite{Deutsch:optical_lattices}. Thus, all degrees of freedom of
the atoms collapse into a single coordinate describing the
position of this fictitous mirror.

The collective back-action of the atoms, as described by what we
may call the ``atom mirror'', strongly affects the light field.
Thus the atoms qualitatively change the potential in which they
move themselves. Furthermore, because of the modifications of the
radiation pressure force, they also influence the mirror motion.

As a consequence of the back-action on the light field the
potential wells in which the atoms sit become narrower and for
certain mirror positions exhibit two local minima per half optical
wavelength, i.e. the atomic position becomes bistable. The
character of the bifurcation points where the single potential
well branches off into two valleys can be modified by changing the
cavity parameters, such as e.g. the reflectivities of the two end
mirrors. In that way it is possible to choose which of the two
valleys is the deeper one and accordingly in which direction the
atoms will move. The two side valleys can also be made equally
deep, in which case the branching point acts like an atomic beam
splitter. We also show that the atomic motion can exhibit
hysteresis, where the atoms move along different paths depending
on the direction of motion of the mirror, giving rise to
additional instabilities of the mirror motion.

In this paper we consider only the adiabatic limit where the
mirror moves slowly compared with the atoms, which in turn
have a much slower response time than the intracavity light field.
Even in that limit the coupled system exhibits a remarkably rich
phenomenology that we illustrate by analyzing and discussing the
modified dipole potential felt by the atoms as well as the
modified motion of the end mirror.

We propose to measure the sideband spectrum of the light
transmitted through the cavity as a quantitative indicator of the
coupled atom-mirror motion, as the coupling of the mirror motion
to the atoms gives rise to unambiguous spectral signatures that
should be within easy experimental reach.

This article is organized as follows: In Section \ref{model} we
describe the model for the coupled atom-mirror-field system and
derive the basic dynamical equations. In section
\ref{adiabaticregime} we consider the dynamics of the mirror with
field and atoms following its motion adiabatically, and discuss
several aspects of the atomic motion.  In section \ref{detection}
we calculate the spectrum of the light transmitted through the
cavity and show that it can serve as a convenient probe of the
coupled mirror-atom motion. Finally, section \ref{outlook} is a
conclusion and outlook.

\section{\label{model} Model}

The system under consideration is shown schematically in Fig.
\ref{setup}. Atoms are trapped in the standing-wave light field of
a Fabry-P{\'e}rot cavity with one of its end mirrors allowed to
move under the effect of radiation pressure and subject to a
harmonic restoring force, $\xi$ being the displacement of the
mirror from its rest position. The cavity length offset $L$ only
gives rise to a shift of the resonances and we will set it to zero
in all that follows.

We measure all lengths in units of the vacuum wavelength of the
light and all times in units of $\Omega_P^{-1}$, where $\Omega_P$
is the oscillator frequency of the unperturbed moving mirror.  The
electrical fields are described in terms of their intensity
integrated over the beam profile, i.e.  in terms of power.

\subsection{Atoms}

Using $\hbar \Omega_P$ as the unit of energy and $\hbar k_0=\hbar
2\pi/\lambda_0$ as the unit of momentum the free hamiltonian of
the atoms takes the form
\begin{equation}
\hat{H}=\int d^3x\hat{\psi}^\dagger (x)
\bigg[-\frac{E_{R}}{4}\nabla^2
+\frac{U_0}{2}\hat{\psi}^\dagger(x)\hat{\psi}(x)\bigg]\hat{\psi}(x),
\label{Hamiltonian}
\end{equation}
where $\hat{\psi}$ and $\hat{\psi}^\dagger$ are the atomic field
annihilation and creation operators. Our theory can easily be
adapted to describe fermions or bosons although at the level of
our present theory there are no significant effects due to the
statistics of the atoms. For the sake of definiteness we
nonetheless assume that the field operators in Eq.
(\ref{Hamiltonian}) satisfy bosonic commutation relations. The
energy
$$
E_R=\frac{(2\hbar k_0)^2}{2M\hbar\Omega_P},
$$
with $M$ the mass of the atoms is the two-photon recoil energy and
$$
U_0=\frac{E_R a}{\lambda_0\pi},
$$
with $a$ the $s$-wave scattering length is the interaction
strength between the atoms describing $s$-wave collisions.

We assume that the atoms can be approximated as two-level atoms
with transition frequency $\omega_a$ and we treat their
interactions with the light field in the dipole and rotating wave
approximations. We consider only one state of polarization of the
light field and describe this component by a scalar field.
Furthermore we assume that the intracavity light field of
frequency $\omega_L$ is far detuned from the atomic transition,
$|\Delta|\equiv |\omega_L-\omega_a|\gg \Omega_R,\gamma$.  Here
$\Omega_R$ is the Rabi frequency of the transition and $\gamma$ is
its linewidth. Then we can adiabatically eliminate the upper state
and, choosing the $z$ axis along the direction of propagation of
the light field, the interaction between light field and atoms is
given by the dipole potential
\begin{equation}
\hat{H}_{\rm int}=g\int d^3x \psi^\dagger(x)\psi(x) |E(z)|^2
|u_\perp (x,y;z)|^2 \label{interaction}
\end{equation}
where
\begin{equation}
g=\frac{2}{\hbar^2c\epsilon_0\lambda_0^2
\Omega_P}\frac{\wp^2}{\Delta},
\end{equation}
where $\wp$ is the electric dipole moment along the polarization
direction of the laser field and the electrical field has been
split into a part varying along the cavity axis $z$ and a
transverse profile $u_\perp(x,y;z)$ assumed to be normalized to
unity,
\begin{equation}
\int dx dy |u_\perp(x,y;z)|^2 =1,
\end{equation}
for every z.
For simplicity we assume in the following that the $z$-dependence
of $u_\perp$ along the atomic cloud can be neglected. We further
assume that the light field is red detuned, $\Delta<0$, so that
the atoms are attracted to the intensity maxima.

The standing wave that forms in the cavity results in an optical
lattice potential. We assume that its wells are very deep, i.e. $g
|E_0|^2\gg E_R$ where $E_0$ is the amplitude of the standing wave.
In the language of condensed matter physics we assume that the
atoms are deeply in the Mott insulator regime \cite{Bloch:MottInsulator1}.
It is then a good
approximation to assume that every site has a specific number of
atoms in the ground state of the harmonic oscillator potential
representing the lattice at that site. Due to the back-action of
the atoms on the light field the distance between the lattice
sites is smaller than $\lambda_0/2$, a point to which we return
shortly.

The interaction between atoms only gives rise to a constant energy
shift. Its influence on the localized wave functions is irrelevant
since all that will be required in the following is that the
atomic density at every site is localized within much better than
an optical wavelength. The order of magnitude of the localization
is given by the oscillator length at a single lattice site
\begin{equation}
a_{{\rm osc},z}=\left(\frac{\pi w^2 E_R}{4 g |E_0|^2 k_0^2}\right)^{1/4}
\label{aosc}
\end{equation}
along the cavity axis and
\begin{equation}
a_{{\rm osc},\perp}=\left(\frac{\pi w^4 E_R}{4 g |E_0|^2}\right)^{1/4}
\end{equation}
in the transverse direction. For typical laser powers and
detunings the oscillator length is roughly ten to a hundred times
smaller than the optical wavelength. The oscillator frequencies
are
\begin{equation}
\omega_{{\rm osc}, z}=\sqrt{\frac{g |E_0|^2 k_0^2 E_R}{\pi w^2 }}
\end{equation}
for oscillations along the cavity axis and
\begin{equation}
\omega_{{\rm osc}, \perp}=\sqrt{\frac{g |E_0|^2 E_R}{\pi w^4 }}
\end{equation}
for oscillations in the transverse direction.  The parameters that
we are interested in are such that the oscillator length in the
longitudinal direction is much smaller than an optical wavelength
and the oscillator length in the transverse direction is much
smaller than the beam waist. Since the beam is typically much
wider than an optical wavelength the atoms can be thought of as
forming a stack of pancake shaped discs.

We consider an optical lattice with $N$ sites occupied by $n$
atoms each. Due to the phase shift suffered by the light upon
propagation through each atomic sheet the lattice sites move
closer together, the self-consistent period of the resulting
lattice being determined later on. If the temperature of the atoms
is far below the oscillator energy, we are justified in neglecting
thermal excitations. If the atoms start in the lattice ground
state, the atomic motion is then reduced to the motion of a
Gaussian wavepacket in a slowly changing harmonic potential, a
situation that is well described by the classical dynamics of the
atomic center of mass.

Because atoms at different sites see the exact same potential at
all times they will be at the same displacement from their
respective local minima given that they started at the bottom of
their respective potential wells. Therefore it is possible to
describe the atomic motion in terms of a single coordinate $z_a$,
the position of each atom modulo the lattice period.

\subsection{Moving mirror}

\begin{figure}
\includegraphics{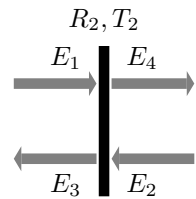}
\caption{Field geometry at movable mirror.}
\label{fieldconfiguration}
\end{figure}

The mirror motion is described by the Newtonian equation of motion
\begin{equation}
\ddot{\xi}+\Gamma \dot{\xi}+\xi=\frac{F_{\rm RP}}{m_P\Omega_P^2\lambda_0}
\equiv \alpha c F_{\rm RP},
\label{newtonmirror}
\end{equation}
where $c$ is the speed of light, $\Gamma^{-1}$ is the mechanical
quality factor of the mirror, $F_{\rm RP}$ is the radiation
pressure force, and
$$
\alpha^{-1}=m_P\Omega_P^2\lambda_0 c
$$
is a power characteristic of the mirror. The electromagnetic field
is treated classically and from Fig. \ref{fieldconfiguration} we
read off that
\begin{eqnarray}
F_{\rm RP}&=&\frac{1}{c}(|E_1|^2+|E_3|^2-|E_4|^2-|E_2|^2)\nonumber\\
    &=&\frac{2|R_2|^2 |E_1|^2}{c}
\label{radiationpressureforce}
\end{eqnarray}
where we have used that in our situation $E_2\equiv0$.

\subsection{Light field}

We now turn to the back-action of the atoms on the light field.
The influence of the cavity mirrors will be considered in the next
section.

It is well known \cite{Meystre:elements_of_quantum_optics} that
for large detuning, optically driving the atoms results in the
polarization
\begin{equation}
P=\frac{-2\wp^2}{\hbar \Delta}\hat{\psi}^\dagger\hat{\psi}Eu_\perp
\end{equation}
leading to the inhomogeneous optical field propagation equation
\begin{equation}
\nabla^2(Eu_{\perp}) + k_0^2 (Eu_\perp)=
\chi |\psi|^2 (Eu_\perp )
\label{waveequation}
\end{equation}
where in our units the polarizability is given by
\begin{equation}
\chi=\frac{2\mu_0\Omega_P}{\hbar}\frac{\wp^2\omega_L^2}{\Delta}.
\end{equation}

In principle, the shape of the transverse profile $u_\perp$ is the
result of the complicated dynamics resulting from the repeated
propagation of the light field through the cavity and its
interaction with the complicatedly structured dielectric
constituted by the atoms. A complete description of these
processes is beyond the scope of this paper. We instead adopt the
point of view that $u_\perp$ has been established in some way and
is assumed to be known. We therefore describe it an effective
manner and eliminate it from the theory.

Specifically, we assume that $u_\perp$ varies slowly with $z$,
$|\partial u_\perp/\partial z |\ll |k_0 u_\perp|$ so that we can
neglect its derivatives with respect to $z$. Then, after dividing
by $Eu_\perp$ to separate the variables, we have from the wave
equation
\begin{equation}
\frac{\partial^2 E(z)}{\partial z^2}+k_0^2 E(z)=\beta^2(z) E(z)
\end{equation}
where $\beta(z)$ is determined for every $z$ by the eigenvalue problem
\begin{equation}
\chi|\psi(x,y,z)|^2 u_\perp(x,y;z) -
\nabla_\perp^2u_\perp(x,y;z)=\beta^2(z) u_\perp(x,y;z).
\label{betasquared}
\end{equation}
In principle this formula solves our problem since it expresses
the propagation equation for $E$ in terms of $\psi$ and $u_\perp$
independently of how the latter was obtained. In order to obtain a
more transparent formula we multiply Eq. \eqref{betasquared} by
$u_\perp^*$, integrate over $x$ and $y$ and after a partial
integration we find
\begin{eqnarray}
\beta^2(z)&=&\chi\int dxdy |u_\perp(x,y,z)|^2|\psi(x,y,z)|^2+\nonumber\\
&&+\int dxdy \left(\nabla_\perp u^*_\perp(x,y;z)\right)\cdot
\left(\nabla_\perp u_\perp(x,y;z)\right)\nonumber\\
&\equiv&\chi \overline{|\psi(z)|^2}+k_\perp^2(z),
\end{eqnarray}
ending up with the intuitively appealing equation
\begin{equation}
\frac{\partial^2 E(z)}{\partial z^2}+(k_0^2-k_\perp^2(z)) E(z)
=\chi \overline{|\psi(z)|^2} E(z), \label{propagationequation}
\end{equation}
where $\overline{|\psi(z)|^2}$ can be interpreted as the density
distribution averaged over the transverse beam profile. In the
following we neglect the reduction of the wave vector $k_\perp$
due to the finite extend of the beam in the transverse direction,
an approximation valid if the beam is not too strongly focused.

For the situation at hand $\overline{|\psi(z)|^2}$ is readily
found, since the wave function of the harmonic oscillator
factorizes into a $z$-dependent part and a transverse part. As
discussed above the atomic density distribution is much narrower
than the beam waist $w$. Hence we find, assuming a Gaussian beam
profile,
\begin{equation}
\overline{|\psi(z)|^2}=\frac{1}{\pi w^2}|\psi_z(z)|^2,
\end{equation}
where $\psi_z(z)$ is the $z$-dependent factor of the oscillator wave
function of the atoms.

\begin{figure}
\includegraphics{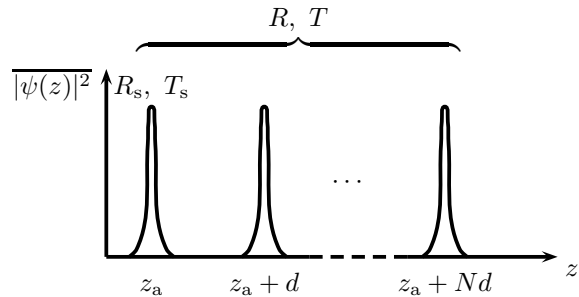}
\caption{Schematic of the Bragg mirror formed by the atoms. Each atomic sheet
has a reflection and transmission coefficient $R_s$ and $T_s$ resulting in a total reflection and transmission coefficient $R$ and $T$.}
\label{latticeschematic}
\end{figure}

Equation \eqref{propagationequation} allows us to determine the
effect of the atoms on the light field. The basic idea is to
replace the entire configuration of atoms by a fictitious ``atom
mirror'' at some position $z_a$ and hence to describe them in
terms of a pair of effective reflection and transmission
coefficients. The period of the self-consistent lattice
automatically fulfills the Bragg condition. Due to the self
consistency requirement this is even true for a lattice with
nonuniform filling, see Ref. \cite{Deutsch:optical_lattices} for
more details. (Parenthetically, this means that we can take into account an
external trapping potential through an appropriate choice of
effective $N$ and $n$.  The thermal motion of the atoms will lower
the reflectivity of the atom mirror and could be taken into
account through a Debye-Waller factor.)
All that remains to do to describe the effect of the
atoms on the light field is to determine the transmission and
reflection coefficient $T$ and $R$ of the atom mirror. A schematic
of the situation is shown in Fig. \ref{latticeschematic}.

Consider first a single atomic sheet. Making use of the fact that
the atoms are localized in the $z$-direction to a much narrower
region than an optical wavelength, we readily find the
transmission and reflection coefficient
\begin{eqnarray}
T_{\rm s}&=&\sqrt{\frac{1}{1+\Lambda^2}}e^{i\phi}\label{ts},\\
R_{\rm s}&=&i\sqrt{\frac{\Lambda^2}{1+\Lambda^2}}e^{i\phi},\label{rs}
\end{eqnarray}
by matching boundary conditions at the atom sheet. Here
\begin{equation}
\phi=-\arctan \Lambda
\label{phaseshift}
\end{equation}
is the phase shift suffered by light upon transmission through the
sheet of atoms and we have introduced the parameter
\begin{equation}
\Lambda=\frac{n \chi}{2\pi k_0 w^2}
\end{equation}
for notational convenience. Equation \eqref{phaseshift} allows us
to determine the self-consistent period of the optical lattice.
Due to the interaction between the light field and the atoms the
period of the optical lattice is reduced to
\begin{equation}
d=\frac{\pi-2\phi}{k_0}.
\end{equation}

\begin{figure}
\includegraphics{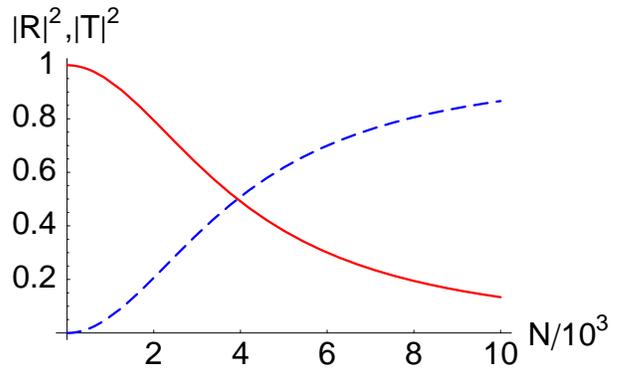}
\caption{(Color online) Reflection (blue dashed line) and
transmission coefficient (red solid line) of an optical lattice
consisting of ten ${}^{87}\rm Rb$ atoms at every site as a
function of the number of lattice sites for $\Delta=10^{9} s^{-1}$
and $w=10 \lambda_0$ as obtained from Eqs.
(\ref{rtot},\ref{ttot}).} \label{trans_refl_tot}
\end{figure}

The entire lattice of atoms can be considered as a periodically
stratified dielectric medium. Its total reflection and
transmission coefficients are readily found using standard methods
of the optics of dielectric films
\cite{Born:principlesofotpics,Deutsch:optical_lattices}. In
essence, one can find the transfer matrix of a unit cell of the
lattice from the transmission and reflection coefficient Eq.
(\ref{rs},\ref{ts}). The transfer matrix of the entire lattice is
then the $N$th power of the transfer matrix of the unit cell. This
can be done in closed form using Tchebychev polynomials
since we neglect scattering losses and therefore the transfer
matrix of the unit cell is unimodular. From the total transfer
matrix the total reflection and transmission coefficient are
extracted as \footnote{Strictly speaking, Eq. \ref{rtot} is the
    the reflection coefficient for a wave propagating in the positive $z$
	direction with the reflection coefficient in the negative $z$
	direction differing by an additional phase $e^{4iN\phi}$. This phase
	is however immaterial for our discussion since it can be absorbed
	in a redefinition of $z_a$. The transmission coefficients for the
	two directions are exactly equal as they have to be for general
	reasons.}
\begin{eqnarray}
R&=&\frac{-i \Lambda N e^{2 i \phi}}{1-i \Lambda N}\label{rtot}\\
T&=&\frac{e^{2 i \phi N}}{1-i \Lambda N}\label{ttot}.
\end{eqnarray}
An example of the transmission and reflection coefficient of such
a lattice of atoms is shown in Fig. \ref{trans_refl_tot} for the
case of ${}^{87}\rm Rb$. From the figure it is apparent that
rather high reflectivities can be obtained in principle. One of
the main experimental difficulties will be to achieve fairly high
filling factors over many sites. Also, the high reflectivities in
Fig. \ref{trans_refl_tot} were obtained by using a small beam
waist and a fairly small detuning of around 100 linewidths. It
should be noted that already fairly moderate reflection
coefficients of the atoms of around $10\%$ give rise to
interesting effects as we discuss later in this paper. S. Slama et al.
have recently reported Bragg reflection coefficients of the order of
$30\%$ \cite{Slama:Multiplereflections}. (Note that
the assumption of a constant and small beam waist is not as bad as
might be expected, as the Gaussian shaped atomic wavepackets at
every site act like lenses that constantly refocus the beam. A
more detailed analysis of the spatial mode structure of the light
field inside the cavity will be presented in a future publication.)

In the following we consider the three representative cases of an
empty cavity, a cavity containing atoms giving rise to an
intermediate reflection coefficient of a few tens of a percent and
atoms with a reflection coefficient close to one.

\section{\label{adiabaticregime} Adiabatic dynamics}

We restrict ourselves to the limit where the motion of the moving
mirror is extremely slow compared to the motion of the atoms which
is in turn very slow compared to the dynamics of the light field.
For this to be true we must have $\Omega_P\ll \omega_{\rm osc,z}$
and $\omega_{\rm osc,z}\ll\kappa$ where $\kappa$ is the linewidth
of the cavity. This is the most relevant case for the typical
systems currently available. The light field then follows the
atoms adiabatically, so that both light field and atoms adjust
instantaneously to any position of the moving mirror. We can then
eliminate the atoms and the light field and express the radiation
pressure force acting on the mirror in terms of the mirror
position alone.

\begin{figure}
\includegraphics{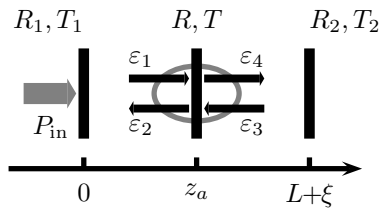}
\caption{Representation of the atoms as a mirror at $z_a$ and
illustration of the steady state boundary conditions at $z_a$.}
\label{boundaryconditions}
\end{figure}

To obtain the effective radiation pressure force in presence of
the atoms we observe that, given positions of the mirror and of the
atoms, the equilibrium light field is determined by the boundary
conditions at the fictitious ``atom mirror,'' see Fig.
\ref{boundaryconditions}.
The steady-state boundary conditions are found to be
\begin{eqnarray}
\epsilon_1&=&T_1 e^{i k_0 z}\sqrt{P_{\rm in}} + e^{2ik_0z}R_1 \epsilon_2\\
\epsilon_2&=&R \epsilon_1+T\epsilon_3\\
\epsilon_3&=&R_2 e^{2ik_0(L+\xi-z)}\epsilon_4\label{eps3}\\
\epsilon_4&=&T\epsilon_1+R\epsilon_3.
\end{eqnarray}
from which we obtain, e.g. for $\epsilon_4$,
\begin{widetext}
\begin{equation}
\epsilon_4=\frac{-e^{ik_0z_a}\sqrt{P_{\rm in}} T T_1}
{RR_1e^{2ik_0z_a}+RR_2e^{2ik_0(L+\xi-z_a)}-
R_1R_2(R^2-T^2)e^{2ik_0(L+\xi)}-1}.
\label{eps4}
\end{equation}
From these fields we find the adiabatic dipole potential felt by
the atoms for a given displacement of the moving mirror as
\begin{equation}
V(z_a;\xi)=g|\epsilon_4+\epsilon_3|^2 =\frac{g P_{{\rm in}} |T_1
T|^2(1+|R_2|^2-2|R_2|\sin2k_0(L+\xi-z_a))}
{\left|RR_1e^{2ik_0z_a}+RR_2e^{2ik_0(L+\xi-z_a)}-
R_1R_2(R^2-T^2)e^{2ik_0(L+\xi)}-1\right|^2},
\end{equation}
\end{widetext}
a result that holds for every lattice site provided that $R_1
\simeq R_2$. Otherwise the depth of the potential wells is
different at each lattice site. \footnote{This should bring about only
minor quantitative changes that are irrelevant for the limit of
the adiabatic motion that we are considering here: We assume from
the outset that the atoms are adiabatically moving in deep
potential wells, i.e. they are always near the bottom of their
respective wells and never see the tops of the potential barriers
separating them.}

The potential $V(z_a;\xi)$ is periodic with period $\lambda_0/2$
in both $z_a$ and $\xi$ direction. In the numerator we have, for
$|R_2|$ close to one, the well known $\cos^2 k_0 (L+\xi-z)$-type
potential. Its minima are moving as the cavity end-mirror is
moving.

\begin{figure}
\includegraphics[width=\columnwidth]{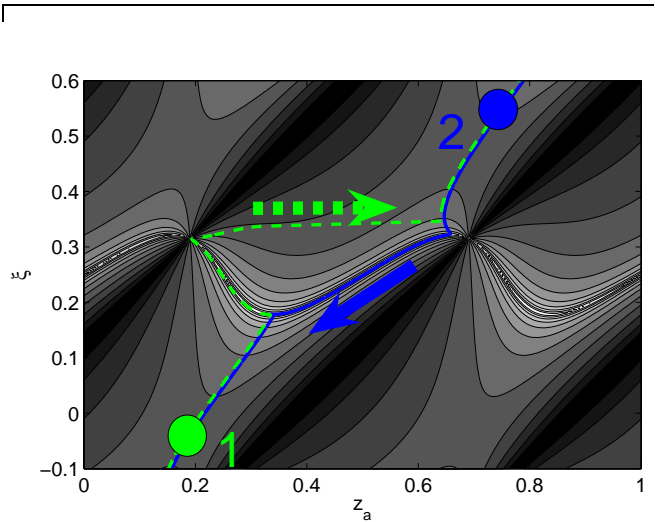}
\caption{(Color online) Dipole potential felt by the atoms for $T_1=0.1$,
    $T_2=0.2$ and relatively high transmissivity of the atomic mirror,
    $T=0.90$. Here, and in the potential plots to follow, lighter regions are
	deeper in potential and the contour lines are drawn at logarithmic
	intervals for clarity and, as in the rest of this paper, all lengths
	are given in units of the wavelength of the injected light field. The
	green dashed line illustrates the path the atoms take when the mirror
	is moving starting at the green dot with the label 1 while the blue
	solid line shows the path the atoms take when the mirror moves back
	down starting at the blue dot labeled 2.}
    \label{vdipolesmallr}
\end{figure}

The effect of the denominator, which gives rise to a series of
resonances, is somewhat less intuitive. If the reflection
coefficient of the atoms is small their back-action on the cavity
field is weak and we see essentially the resonances of an empty
Fabry-P\'erot cavity: Whenever the distance between the two cavity mirrors is
$(m+\frac{1}{2})\lambda_0/2$ with $m$ an integer the field inside the cavity
is very high and accordingly the dipole potential very deep \footnote{The
    additional $\lambda_0/4$ offset comes from our particular choice of phase
	of the reflection and transmission coefficients.}. Fig.
\ref{vdipolesmallr} shows how a small atomic reflectivity slightly distorts
this type of potential. The zeros of the numerator are visible as dark
diagonal lines and the nonzero regions are structured by the back-action of
the atoms on the lightfield.

\begin{figure}
\includegraphics[width=\columnwidth]{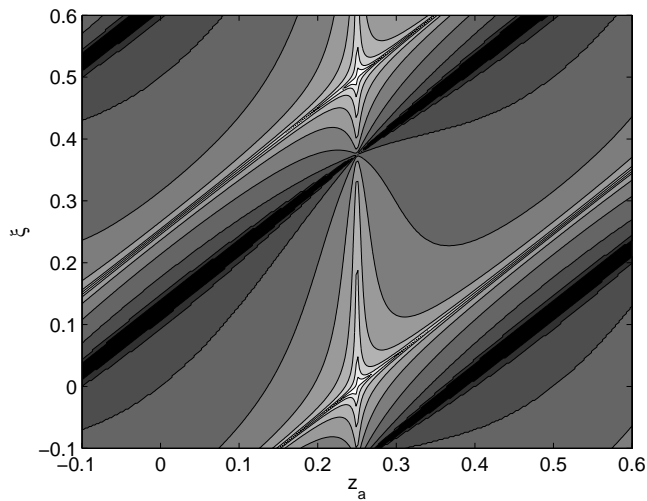}
\caption{Dipole potential for higher atomic reflectivity corresponding to
    $T=0.05$ and $T_1=0.2$, $T_2=0.1$. For this choice of reflectivities of the
    cavity end mirrors the potential well closer to the moving mirror is
    always deeper and, in contrast to Fig. \ref{vdipolesmallr} the atoms
    move on diagonals following the bottom of the potential well for both
    directions of the mirror motion.}
\label{vdipole_movingbig}
\end{figure}

The situation is qualitatively different if the reflection
coefficient of the atoms is high, see Fig \ref{vdipole_movingbig}.
In this case the dipole potential is deep whenever the atoms are
in a position where they form a resonant cavity with one of the
end mirrors.  Interestingly, this typically gives rise to two
potential minima per $\lambda_0/2$, resulting in two distinct
modes of motion for the atoms: They can follow the moving mirror
in the local minimum along diagonals in Fig.
\ref{vdipole_movingbig}, or alternatively they can remain locked
with the fixed mirror in the local minimum following vertical
lines in Fig. \ref{vdipole_movingbig}. These two minima are
actually already present for small atomic reflectivity, but only
for a very narrow range of $\xi$ values near $\xi\approx 0.25$.

\begin{figure}
\includegraphics[width=\columnwidth]{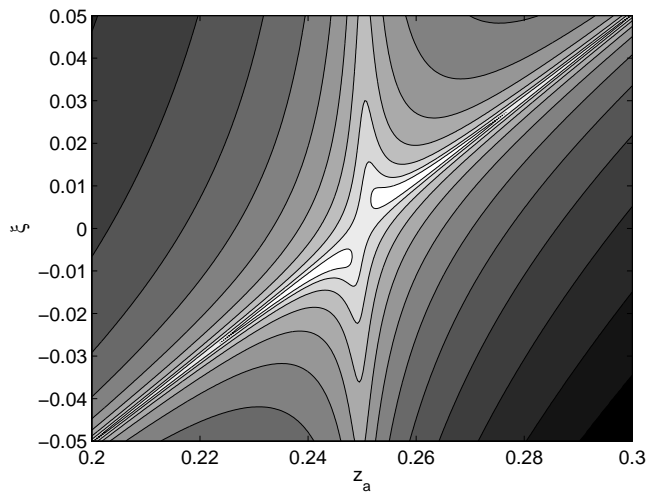}
\caption{Enlargement of the resonance region of Fig.
    \ref{vdipole_movingbig}, showing that
    the potential well corresponding to motion
    with the moving mirror is deeper than the one corresponding to the
    resonance with the fixed mirror, and explaining why the atomic motion is
    locked to the mirror motion. the figure also shows
     the avoided crossing of the dipole
    potential.}
\label{zoomonresonance}
\end{figure}

The relative depth of the two local minima is determined by the
reflectivities of the two end mirrors. The potential well closer
to the mirror of higher reflectivity is deeper. As illustrated in
Fig. \ref{zoomonresonance} the system shows behavior reminiscent
of an ``avoided crossing'' where the two resonances corresponding
to the two modes of motion cross: In positions where the atoms
would be on resonance with both mirrors the potential depth has a
fairly narrow local maximum. The width of the potential barrier
arising this way becomes smaller as the reflectivity of the atomic
mirror increases. If the moving mirror sweeps through such a
resonance, the atoms have to decide which branch of the potential
valley to follow on the other side of the resonance \footnote{The resonances
    act like beam splitters for the atoms and a more exact analysis should
	treat the motion of the atoms near these branching points
	quantum-mechanically.  Also one can easily convince oneself that atoms
	originally in neighboring potential wells can end up in the same
	potential well if they choose to go into different branches at a
	resonance. This provides one with the essential ingredients for an
	interferometer that can in principle operate in a massively parallel
	fashion. This problem will be studied in greater detail in a future
	publication.}.
Since we can decide which branch of the potential well is deeper by choosing
the reflectivities of the end mirrors appropriately we can select the mode of
motion that the atoms will perform.

The adiabatic motion of the atoms is particularly interesting if
the fixed mirror has the higher reflectivity. We discuss it in
detail for the case of small atomic reflectivity.  The
trajectories of the local minima corresponding to the mirror
moving in the positive, resp. negative $\xi$ direction are drawn
as a green dashed line and a blue solid line, respectively, in Fig.
\ref{vdipolesmallr}.  If the moving mirror starts in a region far
off resonance, e.g. at the green dot labeled ``1'' in Fig.
\ref{vdipolesmallr}, there is only one potential minimum corresponding to the
maximum of the numerator in the dipole potential. Moving in the direction of
growing $\xi$ the potential valley eventually reaches a branching point and
the atoms follow it to the left towards the fixed mirror, since the potential
well to the left is deeper if the fixed mirror has the higher reflectivity.
Near the diagonal zero lines of the numerator the potential well becomes
shallower and shallower until finally it ceases to provide a local minimum.
The atoms ``roll'' to the second branch of the potential valley closer to the
moving mirror and follow this local minimum until the system hits the next
resonance, where this story repeats itself.

The atomic motion is totally different if the moving mirror moves
in the negative $\xi$ direction. Then the branch of the potential
closer to the moving mirror provides a local minimum all the way
and the atoms never move to the left branch, see the blue line in
Fig. \ref{vdipolesmallr}.

The motion of the atoms is similar for high atomic reflectivity.
If the mirror is moving in the positive $\xi$ direction the atoms
turn left at every resonance into the vertical potential valley.
Just before reaching the zero potential line they roll to the
right and fall into the diagonal valley. If the mirror moves back
in the negative $\xi$ direction the atoms also turn left at every
resonance and move to the left into the diagonal potential valley
before reaching the zero potential line.

\begin{figure}
\includegraphics[width=\columnwidth]{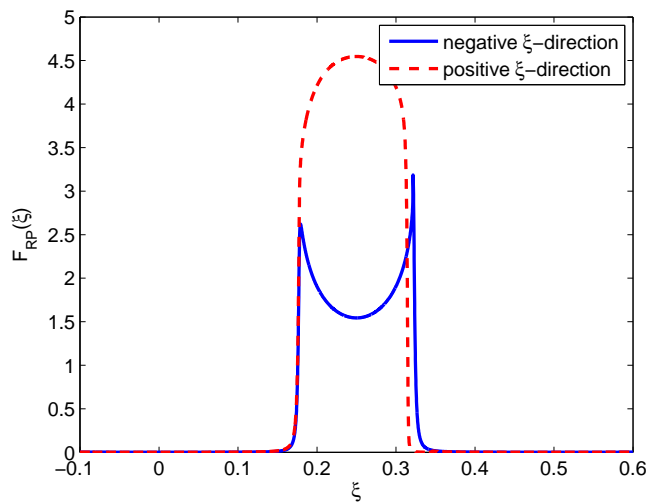}
\caption{(Color Online) Radiation pressure force for the two directions of
    motion of the mirror. The atoms take the paths shown in Fig.
    \ref{vdipolesmallr} for the mirror moving in the positive and negative
    $\xi$-direction, respectively. The difference in area of the two curves
is a measure of the work performed on the moving mirror. All parameters as in
Fig. \ref{vdipolesmallr}.}
\label{hysteresis}
\end{figure}

Interestingly, the radiation pressure force is different for the
two distinct paths the atoms take as the mirror is moving back and
forth, see Fig. \ref{hysteresis}.
Hence the light field performs work on the mirror in every round
trip, and this work can be fairly large.  It should be noted
however that our model is too simplistic to study this feature
quantitatively. First, the assumption of adiabaticity of the
atomic motion is likely to break down near the zeros of the
numerator, where the atoms take a sharp turn to the right. The
potential well becomes extremely shallow in this region and when
the atoms fall into the other branch of the potential, both
situations for which the assumption of adiabaticity of the atomic
motion may not hold. If the atoms have some inertia,
they could overcome the potential wall that prevents them from
going over the potential hill and continue their motion on the
other side. Tunneling through the barrier is another possibility.
For these reasons we concentrate in the following on the case
where the atoms move with the moving mirror, i.e.  where the
reflectivity of the moving mirror is higher than that of the fixed
mirror. For this type of motion the adiabaticity assumption holds
at all times.

\begin{figure}
\includegraphics[width=\columnwidth]{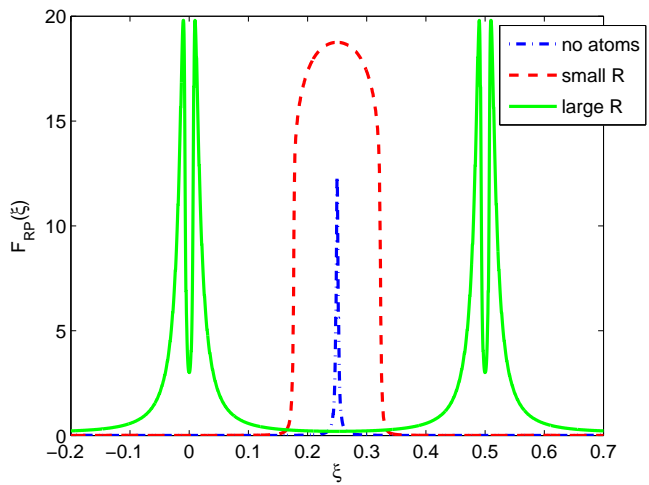}
\caption{(Color online) Radiation pressure force for atomic motion locked to
    the motion of the moving mirror for intermediate atomic transmissivity
    $T=0.9$ (red dashed line), small atomic transmissivity $T=0.05$ (green
        solid line) and an empty cavity (blue dash dotted line) for
    comparison. To ensure that the atoms are following the desired branch
    of the potential we choose $T_1=0.2$, $T_2=0.1$.  The injected laser
    power is $0.01 W$ and $\alpha=10 W^{-1}$.}
\label{radpressure}
\end{figure}

Once the position of the atom mirror $z_a(\xi)$ has been
determined we can find the radiation pressure force by inserting
$z_a(\xi)$ into the expression for $\epsilon_4$. Fig.
\ref{radpressure} shows the three different cases of low atomic
reflection coefficient, high atomic reflection coefficient as well
as an empty cavity for comparison. A small atomic reflectivity
broadens the resonance. Also, the peak radiation pressure force
increases. As the atomic reflectivity increases the resonance
becomes wider and wider and develops a broad dip in the middle.
For very high atomic reflectivities adjacent resonances
``collide'' and give rise to an avoided crossing behavior that
results in the double peak structure of Fig. \ref{radpressure}.
The double resonances in this regime are much wider than the empty
cavity resonances and the peaks are substantially higher. Also the
radiation pressure force never falls off to values as low as for
the empty cavity. The low intensity region between the double
peaks are the limiting shape of the aforementioned dips that
develop in the resonant lineshape.

\begin{figure}
\includegraphics[width=\columnwidth]{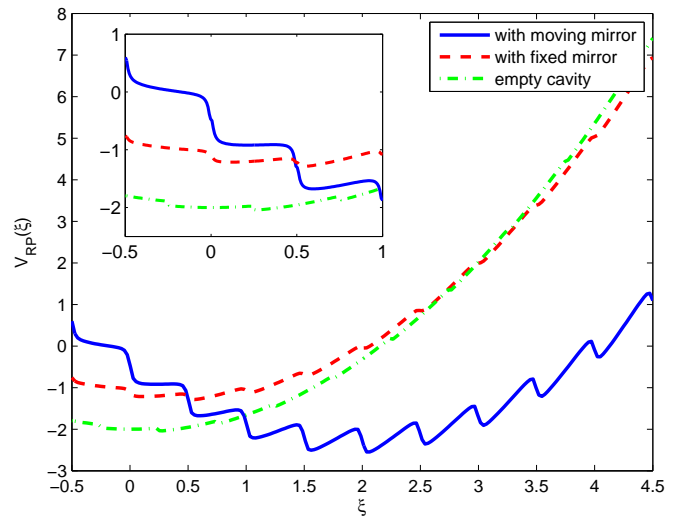}
\caption{(Color online) Radiation pressure potential for atoms
moving with the moving mirror and an empty cavity. All parameters
as in Fig. \ref{radpressure}.} \label{radpotential}
\end{figure}

From the radiation pressure force we find the total potential of
the moving mirror by a simple integration and by adding the
harmonic potential due to the restoring force. Figure
\ref{radpotential} shows the total potentials, again for the three
different cases considered for the radiation pressure force above.
Every time the mirror moves through a resonance the potential
makes a steep step down. The steps are highest if the atoms have
intermediate reflectivity due to the broad and high resonances.
The global minimum of the potential is near $\xi\approx 5$ in this
case. They are smaller for the double peaks for high atomic
reflectivity and smallest for the empty cavity.

\section{\label{detection}Signatures of the coupled motion}

The question of how to detect the coupled motion of the mirror and
the atoms naturally arises. A possible observable is the spectrum
of the light transmitted through the cavity. The amplitude and
phase modulation of the optical field due to the mirror motion are
detectable as sidebands in the spectrum of transmitted light.

To find the spectrum of the transmitted light we integrate the
Newtonian equations of motion of the mirror subject to the
modified radiation pressure force using the velocity Verlet
algorithm. \footnote{The canonical nature of this algorithm is
necessary because we have to integrate the equations of motion
over very long time intervals, typically several ten thousand
cycles. At the same time we have to integrate with high temporal
resolution because the radiation pressure force is so sharply
peaked. The position of the atoms is determined at every time step
using Brent's method \cite{Press:numericalrecipes} in an interval
centered at the atomic position of the previous time step and with
a width that depends on the velocity of the mirror.  This way we
ensure that the atoms follow a continuous path along a potential
valley without jumps to different valleys. The transmitted field
at every time is easily found from $\epsilon_4$. The side band
spectrum is found by numerically fourier transforming the field.}

\subsection{Large amplitude oscillations}

First we analyze what happens as the mirror undergoes large
amplitude oscillations such that in every period it sweeps over
several resonances. \footnote{To avoid too large velocities at the
bottom of the potential, which would make an accurate numerical
integration more difficult, we choose the initial conditions such
that the mirror swings through roughly eight resonances. This
number is not exact since the turning point on the left is
typically at a resonance.} We consider again the three cases of
atoms with high reflectivity, intermediate reflectivity and an
empty cavity. The cavity and atomic parameters are the same as
before and as for the quality factor of the cavity we choose
$\Gamma^{-1}=10^4$.

\begin{figure}
\includegraphics[width=\columnwidth]{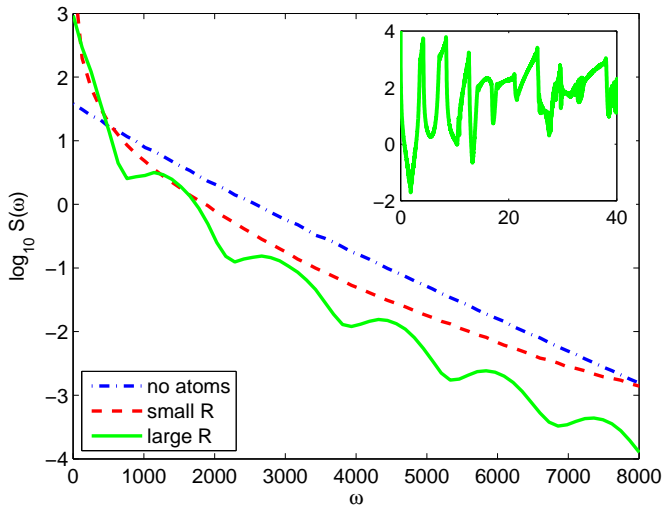}
\caption{(Color online) Sidebands of the transmitted light for
atoms moving with the mirror, staying with the fixed mirror and
for an empty cavity.} \label{spectra}
\end{figure}

Figure \ref{spectra} shows the resulting sideband spectra. For
clarity each data point has been obtained by averaging over 1000
frequency bins. The spectrum of the empty cavity falls off nearly
exponentially, a consequence of the near-Lorentzian lineshape of
the resonances and the fact that the mirror moves through each
resonance at nearly constant velocity.

The widths of the transmission spectra are comparable in all three
cases, confirming that the widths of the resonances are of the
same order. However the spectra acquire a peak near $\omega=0$ for
nonzero atomic reflectivities, a consequence of the larger
resonance widths in these cases; in addition the light field never
falls off to zero, in contrast to the empty cavity case where it
does so to a good approximation for high-reflectivity mirrors.
Especially for high atomic reflectivity the cavity always leaks a
significant amount of light that gives rise to an enhanced DC
component in the transmission spectrum.

The fringes that appear in the case of high atomic reflectivity
can be understood in the following way: two light pulses exit the
cavity whenever the mirror passes through a resonance, see Fig.
10, and they give rise to a double-slit diffraction pattern in
frequency space. Their spacing corresponds to the inverse of the
time it takes the mirror to travel from the first peak to the
second peak of the double resonance.

The inset of Fig. \ref{spectra} shows the details of the low
frequency part of the spectrum for the case of high atomic
reflectivity. (That low frequency part is similar in all
cases.) It consists of a peak at the frequency corresponding to
the time it takes the mirror to move from one resonance to the
next and of a series of additional peaks at each
harmonic. These strong components are a result of the highly
nonlinear potential. The fact that the mirror moves at different
speeds between different resonances gives rise to a broadening of
the harmonics into bands whose width increases with growing
harmonic number to finally overlap. Effects due to the damping of
the mirror motion cannot be resolved because of this ``inhomogeneous
broadening''.

\subsection{Small amplitude oscillations}

In our units the average thermal energy of the mirror at room temperature is
$T\approx 10^{-5}$ . Comparing with the energy scale in Fig.
\ref{radpotential} we see that the mirror can undergo many oscillations in
each one of the sawtooth-shaped local potential minima before it hops to a
different local minimum due to thermal activation or some other perturbation.
Thus it is an interesting question to determine the oscillation frequency at
every local minimum for small amplitude oscillations.

\begin{figure}
\includegraphics[width=\columnwidth]{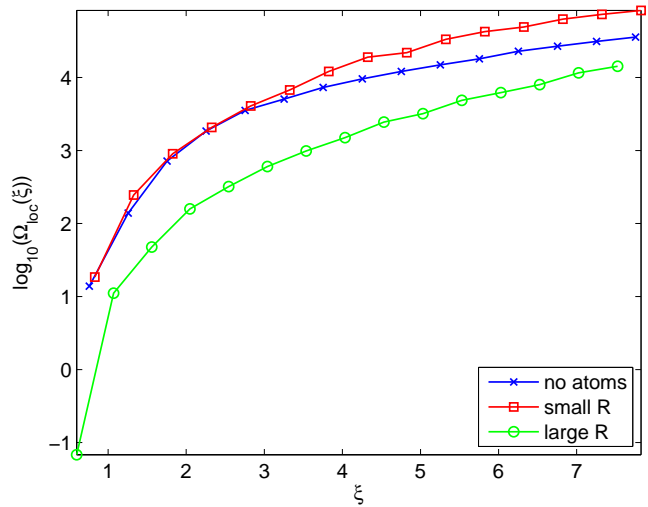}
\caption{(Color online) Approximate harmonic frequencies near the local minima
    for atoms with high reflectivity, intermediate reflectivity and an empty
    cavity. All parameters as in Fig \ref{spectra}.}
\label{locfrequencies}
\end{figure}

Figure \ref{locfrequencies} shows these frequencies for the three
cases already discussed. For high atomic reflectivity, and in
contrast to the intermediate reflectivity and empty cavity cases,
the oscillation frequency near $\xi=0$ is much lower than the
restoring frequency of the mirror. This is because the resonances
have wide wings in this case and, if added to the harmonic
oscillator potential of the mirror, they can compensate the
harmonic restoring force almost completely over a large range of
displacements $\xi$ and give rise to an extremely flat potential
near the local minimum. In the other two cases the onset of the
radiation pressure force near the resonances is very abrupt and
the frequency of the mirror motion can never be significantly
smaller than the oscillator frequency. At higher displacements
from the mirror rest positions the stronger resonances found with
atoms inside the cavity give rise to stronger confinement and lead
to resonance frequencies that eventually become higher than in the
empty cavity case. Note that for the empty cavity cases we could
find only a few more local minima beyond $\xi=8$ and these could
not be reasonably approximated by a harmonic potential.

\section{\label{outlook}Conclusion and Outlook}

We have developed a model that describes the coupled motion of a
cavity end-mirror and atoms trapped in the standing-wave
intracavity field. The key result is that the resulting model is
simple enough to allow for a fairly straightforward analysis. The
self-consistent electric field inside the cavity containing the
atoms plays a central role since from this field we find the
modified dipole potential in which the atoms move as well as the
modified radiation pressure force acting on the moving mirror. We
find that as a result of the strong coupling between light field
and atoms due to the collective back-action of the atoms the
dipole potential changes qualitatively, the most notable feature
being that the atomic position becomes bistable. The resonances of
the cavity are broadened and develop a double peak structure in
the limit of high atomic reflectivity. We have studied the
dynamics of the mirror subject to the modified radiation pressure
force in the adiabatic limit where the atomic motion is much
faster than the motion of the mirror and the light field is slaved
to both atoms and mirror. We have shown that the sidebands of the
light transmitted through the cavity can serve as an indicator of
the coupled motion of the mirror and the atoms.

Many open questions remain to be studied. First of all a more
careful analysis of the spatial mode structure inside the cavity
in the presence of the atoms is desirable. Such an analysis could
start from the Maxwell-Bloch equations and should include
scattering losses due to spontaneous emission from the atomic
excited state. It would yield more accurate values for the
effective reflection coefficient of the atoms and the modified
dipole potential, especially if $R_1$ and $R_2$ are very
different.

One should also consider different adiabaticity regimes, or
abandon the assumption of adiabaticity altogether. The situation
where the atomic motion is roughly as fast as the mirror motion
and the atom and mirror influence each other on an equal footing
is particularly interesting, and so is the case where the
retardation of the light field is taken into account. Already in
an empty cavity the retardation of the light field gives rise to
many interesting phenomena such as instabilities, self sustained
oscillations and chaotic behavior. It will be interesting to see
how all these effects are influenced by the presence of the atoms.

A related question of practical significance is what happens if
the atoms are subject to a friction force. Atoms can be
efficiently cooled to very low temperatures with laser cooling and
one could hope that by means of the coupling to the mirror through
the cavity field one should be able to also cool the mirror.
Loosely speaking, if the atoms are subject to a friction force,
they lag slightly behind the position at which they ought to be
according to the position of the mirror. This retardation has been
seen to lead to the damping of the mirror motion in preliminary
results, but these calculations need to be carried out in much
more detail, especially since the momentum kicks in laser cooling,
and hence the fluctuations in the atomic position, are substantial
and cannot be neglected. They can be taken into account e.g. using
a master equations approach.

Finally, one needs to address the situation where all constituents
of our model are quantized. Quantization of the atomic motion
should allow one to study whether the resonances can be used as
beam splitters, and the properties of the resulting
interferometer. Also, a detailed understanding of the role of
damping of the atomic motion requires a proper quantum treatment.
The quantization of the mirror motion is called for in view of the
the experimental goal to cool it to its quantum mechanical ground
state. Finally, a quantum treatment of the light field, and
especially of its fluctuations, is interesting because of the
powerful methods to measure the fluctuations of the
electromagnetic field. If they are linked to fluctuations of the
atomic position or the mirror position, as must be expected, one
could then use the light field as a powerful diagnostic tool for
the study of these mechanical fluctuations.

\section{Acknowledgements}

This work is supported in part by the US Office of Naval Research,
by the National Science Foundation, by the US Army Research
Office, by the Joint Services Optics Program, and by the National
Aeronautics and Space Administration.

\bibliography{mybibliography}
\end{document}